\documentclass[journal]{IEEEtran}

\usepackage{cite}
\usepackage{microtype}
\usepackage{amsmath,amssymb}
\usepackage{graphicx}
\usepackage{booktabs}
\usepackage{array}
\usepackage[dvipsnames,svgnames]{xcolor}
\usepackage[hyphens]{url}
\usepackage[hidelinks]{hyperref}

\providecommand{\Description}[1]{}

\hypersetup{
  pdftitle={What the Middleman Delivers: Answer Behavior, Tool Permissions, and Billed Usage in LLM API Relays},
  pdfauthor={Xiaodong Wu, Zhimin Zhao, Qi Li, Chaklam Cheong, Xiangman Li, Jianbing Ni},
  pdfsubject={LLM API relay measurement},
  pdfkeywords={LLM API relays, security, measurement, tool permissions, billed usage}
}

\begin{document}
\bstctlcite{IEEEexample:BSTcontrol}

\title{What the Middleman Delivers:\\
Answer Behavior, Tool Permissions,\\
and Billed Usage in LLM API Relays}

\author{Xiaodong Wu,~\IEEEmembership{Student~Member,~IEEE,}
Zhimin Zhao, Qi Li, Chaklam Cheong,\\
Xiangman Li,~\IEEEmembership{Student~Member,~IEEE,}
and Jianbing Ni,~\IEEEmembership{Senior~Member,~IEEE}%
\thanks{All authors are with Queen's University, Kingston, Ontario, Canada
(e-mail: xiaodong.wu@queensu.ca, zhimin.zhao@queensu.ca, qi.li@queensu.ca,
chaklam.cheong@queensu.ca, xiangman.li@queensu.ca, jianbing.ni@queensu.ca).}%
\thanks{Xiaodong Wu, Zhimin Zhao, and Qi Li contributed equally to this work.}\thanks{Corresponding author: Jianbing Ni (jianbing.ni@queensu.ca).}}

\pagestyle{plain}

\maketitle
\thispagestyle{plain}

\begin{abstract}
Third-party application programming interface (API) relays offer access to
large language models (LLMs), often at lower prices. Users depend on these
services for model quality, instruction compliance, and accurate billing.
However, existing studies leave unclear whether relays meet
these expectations together. We test 20
relays offering GPT-6 Astra and Claude Opus 5. First, we compare answers to
knowledge questions with batch-matched official responses. In our first-round
sample, the analysis flags 7/10 low-price and 2/10 same- or higher-price relays
for behavioral differences. On retesting 35 pairs, seven gain or lose a flag
under the original rules, and six under matched correction families. Because answer comparisons
cannot reveal every delivery problem, we examine returned content and tool
permissions. No injection is confirmed in 3,332 screened responses to a fixed,
non-adversarial code and setup workload, yet four relays return tool calls despite instructions disabling
them in the first round. Finally, to connect reported usage to actual costs,
we compare input-token counts with official references and check
balances. For identical requests, five relays report 4,200 to 60,000 more input
tokens than the official API. Where balances are available (six services), charges follow
reported usage at listed prices, including extra reported input tokens at one service. Our study shows
that advertised models and prices are insufficient to judge relay reliability.
It identifies unexplained billed token quantities and tool-permission violations as
risks and shows why service evaluations need repeated measurements.
\end{abstract}

\noindent\textbf{Impact Statement.}
LLM API relays make advanced models easier to access, but they also place an
operator-controlled delivery path between an application and the service it
selected. The resulting contract covers more than a model label: it includes
response behavior, tool permissions, response fields, and reported usage that
may determine charges. Our measurement makes these client-visible properties
testable through official comparisons, response review, and permission checks.
It supports enforcing tool permissions in client applications, recording
observed discrepancies, and making purchasing decisions with explicit
uncertainty about hidden routing and future behavior. The findings do not
identify a backend or accuse a provider of malicious conduct. Instead, they
give developers and service operators a documented way to compare delivery
responsibly, including negative findings, without treating price or model labels as a guarantee.

\begin{IEEEkeywords}
Large language models, application programming interface security, man-in-the-middle attacks, model substitution, response integrity, supply chain security
\end{IEEEkeywords}

\section{Introduction}
\label{sec:intro}
\begingroup
\clubpenalty=10000

Third-party large language model (LLM) relays let users access models from
multiple providers through a single application programming interface (API).
Users turn to these services for lower prices, convenient payment options, or
access to models that are unavailable in their region~\cite{zhang2026shadow}.
Instead of sending requests directly to a model provider, users send them to
the relay, which chooses where and how to forward them. Even when each
connection is encrypted, a relay that handles plaintext can read and change
the content passing through it~\cite{xie2026aegis}.
This control leaves users unable to verify important parts of the service
from an answer alone. They need the relay to preserve instructions, respect
tool permissions, and report usage accurately. Yet a plausible answer can
still be accompanied by a tool call the user disabled. Likewise, a charge can
match the listed token price while charging for unexplained input-token quantities.
Without seeing what the relay sends upstream, users cannot resolve these
questions from a model label or an invoice. Evaluating behavior, permissions,
and usage together is therefore necessary to understand what users receive
and pay for.

\begin{figure*}[t]
  \centering
  \includegraphics[width=0.9\textwidth]{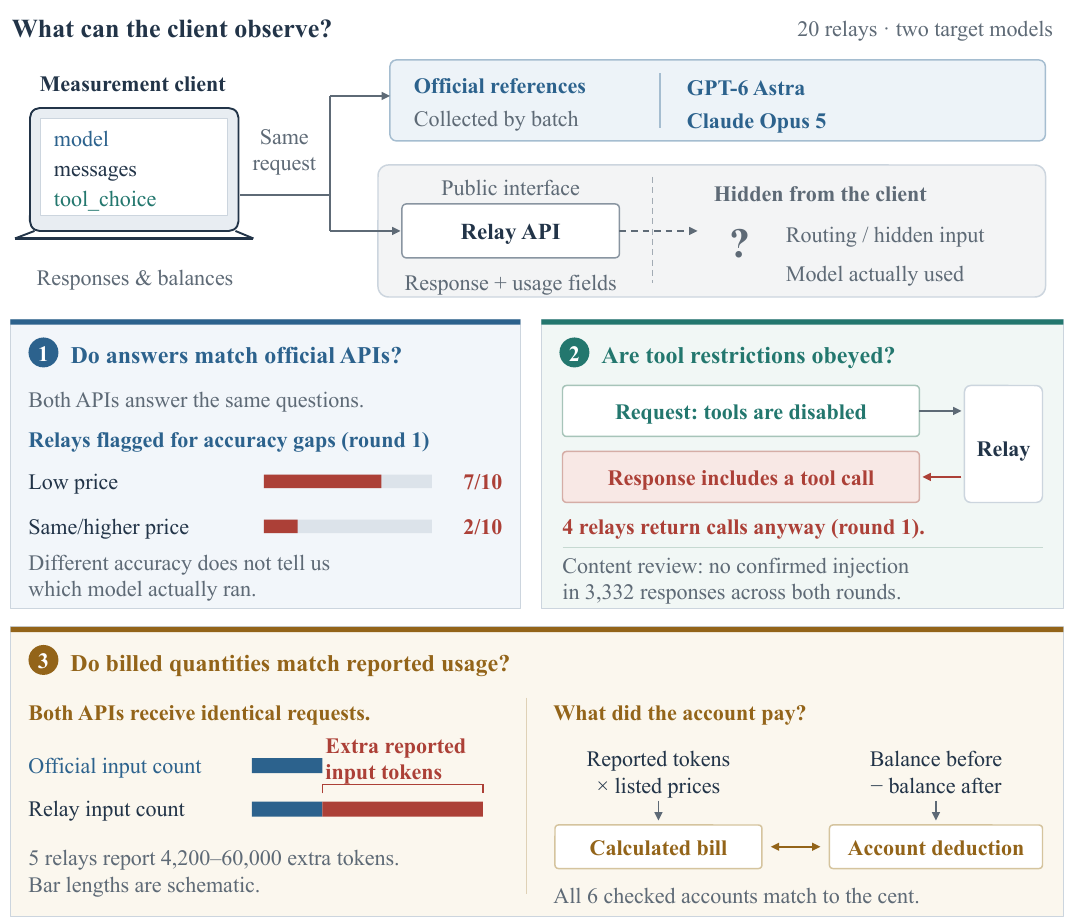}
  \caption{Measurement setup and comparisons of relay answers, tool permissions,
  and billed usage.}
  \Description{The upper scene shows identical requests sent to official
  references and relay APIs, with the relay's internal processing hidden from
  the client. Three groups explain the measurements. First, official and relay
  APIs answer the same questions. Bars show accuracy differences at seven of
  ten low-price and two of ten same- or higher-price relays in round one.
  Different accuracy does not identify which model ran. Second, a request
  disables tools, but the relay returns a response containing a tool call.
  Four relays do this in round one; no calls are executed. Separate content
  review confirms no injection in 3,332 screened responses across both rounds.
  Third, identical requests receive different reported input counts. A
  schematic bar separates the official count from the extra count reported by
  the relay. Five relays report approximately 4,200 to 60,000 extra input
  tokens, summarized by second-round medians. Alongside this, reported token
  counts and listed prices produce a calculated bill, while before-minus-after
  balances yield the actual account deduction. These amounts agree to the
  cent at all six checked services.}
  \label{fig:scenario}
\end{figure*}

Existing research addresses parts of this problem through service measurements
and model verification. A study of shadow APIs identifies 17 services and
evaluates three for utility,
safety, and model identity~\cite{zhang2026shadow}. GateScope measures ten
gateways for response consistency, multi-turn memory, billing, and
latency~\cite{lin2026gatescope}. For model verification, Knowledge Boundary as
Fingerprint (KBF) uses factual recall near a model's knowledge boundary to
compare API endpoints~\cite{fang2026kbf}. Together, these studies provide
methods and measurements for examining relay behavior. However, a model check
cannot establish whether a relay respects tool restrictions, and checking
listed prices cannot verify the source of billed input-token quantities. It remains
unclear which of these problems occur together within a service and whether
they persist.

To address these questions, we study 20 relays offering GPT-6 Astra and Claude
Opus 5. We found candidates through recommendations from LLMs, community
lists, and web searches, then selected services that met our access and
payment requirements. The selection was not random. Ten services advertise
prices below the official APIs, and ten advertise the same or higher prices.
We first compare relay answers with batch-matched official responses to the same questions. Building on KBF's
knowledge-boundary probes, we select questions that distinguish the targets
from candidate substitutes and score answer correctness. Controlled mixtures
of official and substitute responses then show when the tests can detect a
difference. Because answer accuracy alone cannot reveal every problem, we also
screen returned content and check whether tool calls comply with the request.
To connect usage reports to actual charges, we compare input-token counts for
identical requests and check the costs against account balance changes.
Figure~\ref{fig:scenario} outlines the tests.

A single measurement, however, cannot show whether an observed problem
persists. We therefore repeat the tests 11--18 days later for 35 of the 36
relay--model pairs measured in the first round. Fresh official responses
provide a reference for each round, helping distinguish changes in relay
behavior from changes in the official service. Tracking the same pairs also
lets us separate recurring discrepancies from those that appear or disappear
between rounds. Together, these comparisons allow users to detect problems
through public APIs and account records, without relay cooperation. They
measure what the client receives, but do not identify the model.

Our measurements reveal three main findings. First, in the first-round
sample, the analysis flags 7/10 low-price and 2/10 same- or higher-price relays
for behavioral differences from the official reference for at least one target model
($p=0.07$ for the price-group comparison). Six services are flagged in
both rounds, but 7/35 relay--model pairs gain or lose a flag under the original rules. Second, no injection is confirmed
in 3,332 screened responses, yet four relays return tool calls despite
explicit restrictions in the first round. These calls are not executed.
Third, five relays report 4,200 to 60,000 more input tokens than the official
API for identical requests. At all six services with accessible balances,
deductions match reported usage at listed prices to the cent, including the
reported surplus at
R11. Thus, listed-rate billing does not establish whether extra reported
tokens represent added input or altered accounting.

Our contributions are:
\begin{itemize}
  \item We document both recurring and changing behavioral differences across
        20 relays, showing why reliability assessments need repeated comparisons
        with official references
        (Sections~\ref{sec:results:rq2} and~\ref{sec:results:rq4}).
  \item We demonstrate that extra reported input-token quantities enter R11's bill,
        showing why billing checks must verify reported quantities as well as
        listed prices (Section~\ref{sec:results:rq5}).
  \item We observe returned tool calls that contradict explicit restrictions
        in the first round (none in the second), despite no confirmed injection
        in the screened responses, showing why tool permissions need checks
        beyond answer quality and content screening
        (Sections~\ref{sec:results:rq3} and~\ref{sec:results:rq5}).
  \item We provide a documented procedure for evaluating relays without
        internal access. Its controlled-mixture tests, reviewed labels,
        explicit scoring rules, and anonymized summaries explain
        the checks and what the tests can and cannot detect
        (Sections~\ref{sec:methodology:controls} and~\ref{sec:ethics:artifacts}).
\end{itemize}
\par\endgroup

\section{Background and Threat Model}
\label{sec:background}

\subsection{How Relay Services Work}
\label{sec:bg:path}

A relay sits between a client application and a model provider. The client
sends the relay a model name, messages, instructions, tool permissions, and
generation settings. The relay chooses an upstream service, forwards the
request, and returns an answer with fields such as the model name and token
usage. Users pay the relay and can inspect its responses and account records,
but usually cannot inspect what it forwards upstream.

We study relays that can read the request contents. Transport layer security
(TLS) encrypts each connection, but the relay decrypts the client's request
before forwarding it over another connection~\cite{xie2026aegis}. It can
therefore change instructions, choose another model, or edit the response.
If several intermediaries handle a request, the final response may reveal a
problem without showing which intermediary caused it.

Some changes are part of normal service operation. A relay may convert API
formats, cache responses, or use a fallback that the user has accepted.
We compare answers with the official reference, check tool permissions,
and reconcile usage with listed rates. This requires more than reading
the returned model name or invoice, since the relay controls both. Our tests
use registered relay and official accounts without access to the relay's
internal systems or provider accounts.
An official-reference difference alone does not establish a breach of service
terms: disclosed wrappers, fallbacks, or overhead may be part of the purchased
service. Explicit tool suppression supplies a directly testable request constraint.

\subsection{What We Measure and What We Cannot Observe}
\label{sec:bg:threat}

We assume that a relay operator can read and modify requests and responses,
choose the upstream model, change generation settings, and report different
model names or token counts. Its behavior may vary by account, prompt, or
whether it recognizes a test. We assume that the client, official reference
accounts, and encrypted connections are not compromised. Under these
assumptions, we examine three parts of the service using public interfaces.

First, we compare relay answers with official answers to the same questions.
A relay may use a different model for all requests or only some of them.
The controlled mixtures test how this affects our ability to detect an
accuracy difference. However, changed instructions, settings, or cached
responses can also affect accuracy, so the comparison cannot identify which
model actually ran. Reproducing this comparison requires official API access.

Second, we inspect returned content and tool calls. A relay can change an
answer even when it uses the requested model. We screen code and setup
responses for potentially harmful packages, commands, and destinations, then
review the selected responses in context. We separately check whether
returned tool calls obey the request's permissions. These checks distinguish
problems in the response from differences in answer accuracy. Our tests do not execute returned commands or tool calls.

Third, we compare reported usage and charges. Identical requests sent to a
relay and an official API provide a reference for input-token counts.
Where account balances are available, we also compare deductions with the
cost calculated from reported usage and listed prices. These comparisons
show what users are charged for, but cannot reveal the content of any hidden
input or determine why the token counts differ.

Table~\ref{tab:threat-model} summarizes these measurements. They detect
problems visible to clients without requiring the relay's cooperation.
They do not verify the relay's internal processing.

\begin{table}[t]
\caption{Relay capabilities and the observable outcomes covered by this study.}
\label{tab:threat-model}
\centering\small
\setlength{\tabcolsep}{3pt}
\begin{tabular}{@{}>{\raggedright\arraybackslash}p{.23\columnwidth}
>{\raggedright\arraybackslash}p{.35\columnwidth}
>{\raggedright\arraybackslash}p{.37\columnwidth}@{}}
\toprule
Threat & Relay capability & Observable results \\
\midrule
Model substitution & Change model or settings & Answer accuracy comparisons (RQ1, RQ2) \\
\addlinespace
Content and tool changes & Edit instructions, outputs, or tool calls & Content review and permission checks (RQ3, RQ5) \\
\addlinespace
Usage and charge changes & Change reported token counts or charges & Token counts and balance changes (RQ5) \\
\midrule
\multicolumn{3}{@{}l@{}}{\textit{Outside the measurement scope}} \\
\addlinespace[2pt]
\multicolumn{3}{@{}l@{}}{Conversation retention: store messages} \\
\multicolumn{3}{@{}l@{}}{Credential misuse: reuse or expose API keys} \\
\multicolumn{3}{@{}l@{}}{Gateway compromise: exploit relay software} \\
\bottomrule
\end{tabular}
\end{table}

Some risks leave no visible change in the response. A relay can store
conversations while returning correct answers. Privacy-policy studies examine
what model developers claim to retain~\cite{king2025privacy}. CodeCloak reduces code
exposure in prompts~\cite{finkman2024codecloak}, and confidential-computing
systems protect inputs within a trusted runtime~\cite{li2024confidential}.
Our response comparisons cannot establish whether a relay stores or deletes
messages. Likewise, our standalone requests may miss changes triggered only
by longer conversations, credentials, or tool execution.

We also do not test stored API keys or the relay's software. Leaked keys and
unsafe integrations have been studied in mobile traffic and agent
skills~\cite{gao2026mindkey,chen2026skills}. A LiteLLM vulnerability shows how
relay software can expose managed credentials~\cite{cve2026litellm}, even
when the operator intends to forward requests correctly. These are relevant
risks, but we do not determine whether the measured services have them.

\section{Related Work}
\label{sec:related}

\begin{table*}[t]
\caption{Focus and methods of model checks and relay measurements.}
\label{tab:related}
\centering\small
\setlength{\tabcolsep}{4pt}
\begin{tabular}{@{}>{\raggedright\arraybackslash}p{.23\textwidth}
>{\raggedright\arraybackslash}p{.29\textwidth}
>{\raggedright\arraybackslash}p{.43\textwidth}@{}}
\toprule
Approach & What it checks & Method and access \\
\midrule
Output fingerprints~\cite{pasquini2025llmmap,bruckner2026onetoken} &
Model identity & Match outputs to known model profiles \\
\addlinespace[2pt]
KBF~\cite{fang2026kbf} &
Model behavior & Match stable correct and incorrect answers \\
\addlinespace[2pt]
Equality and rank tests~\cite{gao2025equality,zhu2025rut} &
Output-distribution equality & Compare completion samples; rank tests need local reference scoring \\
\addlinespace[2pt]
Computation verification~\cite{sun2024svip,pal2026privacy,jin2026aftune} &
Model execution & Require server participation or trusted execution \\
\addlinespace[2pt]
GateScope~\cite{lin2026gatescope} &
Consistency, memory, billing, latency & Use behavioral probes and local token estimates \\
\addlinespace[2pt]
Shadow-API study~\cite{zhang2026shadow} &
Utility, safety, model claims & Compare benchmarks and fingerprints with official models \\
\addlinespace[2pt]
Agent-router study~\cite{liu2026youragent} &
Malicious edits and credential misuse & Test agent workflows with tools and credentials \\
\midrule
This study & Accuracy, permissions, billed input &
Use official API references and balance checks across price groups and two rounds \\
\bottomrule
\end{tabular}
\end{table*}

\subsection{Checking Models through API Outputs}

Model checks use API responses to identify a model or detect a difference
from a reference. LLMmap identifies models from characteristic
responses~\cite{pasquini2025llmmap}, while single-token fingerprints use
response frequencies from short random-generation prompts~\cite{bruckner2026onetoken}.
IRIS compares responses with profiles of known backends to detect model
substitution, including cases where only some requests use another
model~\cite{zhang2026iris}. Their coverage depends on the reference models.
Statistical tests compare
completion distributions~\cite{gao2025equality}. Rank-based uniformity testing,
for example, scores remote completions with a locally accessible reference
model and tests their ranks~\cite{zhu2025rut}. It requires local reference
scoring, but does not require log probabilities from the tested API. Our
comparison uses answers from hosted official APIs and measures accuracy
differences, without identifying a replacement model. Watermarking offers
another approach: Dasgupta et al. use language-model-generated instructions
to embed detectable patterns in responses~\cite{dasgupta2026watermarking}.
This requires control over the instructions used to generate marked outputs.
Our comparisons require no prearranged watermark.

Knowledge Boundary as Fingerprint (KBF) provides the foundation for our
question design. It uses factual recall near a model's knowledge boundary,
including stable correct and incorrect answers, to distinguish API
behavior~\cite{fang2026kbf}. Building on this idea, we select questions that
the official target answers correctly and multiple cheaper candidates miss.
We then compare answer correctness, using our own selection and decision
rules. We use these probes to examine accuracy together with content, tool permissions,
and billed usage
across price groups and repeated measurements. The controlled-mixture tests
calibrate our implementation and do not establish superiority to KBF.

A further challenge is that services may change over time or replace only
some responses. Border-input tracking detects changes from a baseline without
establishing the baseline model's identity~\cite{chauvin2026b3it}, and studies
of partial substitution examine the difficulty of detecting mixed
responses~\cite{cai2025paying}. Token-DiFR compares tokens with a trusted
reference implementation, sharing sampling seeds for stochastic
decoding~\cite{karvonen2025difr}. Nasery et al. study adversarial evasion
of fingerprint checks~\cite{nasery2025robust}, while GhostPrint fine-tunes
a weaker model to mimic stronger-model fingerprints~\cite{zhang2026ghostprint}.
Our mixtures test replacement fractions, and repeated collections track
behavioral changes. Neither tests adversarial evasion or imitation. Table~\ref{tab:related} compares the related approaches.

\subsection{Verifying Computation and Delivery}

Verification protocols can check model execution when the server
participates. SVIP uses hidden representations and a secret verification
task~\cite{sun2024svip}, while protocols based on privacy-preserving inference
add checks to participating services~\cite{pal2026privacy}. AFTUNE verifies sampled
execution-trace blocks inside a trusted execution environment
(TEE)~\cite{jin2026aftune}. These methods require access beyond the final text
response and do not by themselves protect an answer from later changes along the
delivery path. Signed responses address this separate problem by binding
specified execution fields to a trusted provider~\cite{luo2026alignment}.
Attestation extensions also support request binding and records of output
transformations~\cite{guan2026aex}. Their protection depends on which objects
are signed and which issuers the client trusts.

Other systems protect the forwarding process itself. AEGIS places forwarding
inside an enclave that the client verifies before sending
plaintext~\cite{xie2026aegis}. Gateway designs also separate operator control
from an attested runtime and routing records~\cite{wang2026evidence}.
These approaches require support from a provider or forwarding component.
Our checks use existing public APIs and account records to reveal discrepancies.
They cannot authenticate the model or the internal delivery path.

\subsection{Relay Measurements and Practical Monitoring}

Zhang et al. organize LLM applications by user roles and task types to guide
model selection~\cite{zhang2026applications}. Relay measurements address the
next question: whether the service delivers the expected behavior.
GateScope measures gateways for response consistency, multi-turn memory,
billing, and latency~\cite{lin2026gatescope}. Shadow-API research examines
utility, safety, and model verification, including their consequences for
reproducibility~\cite{zhang2026shadow}. Agent-router studies extend this
question to workflows involving tools and credentials, finding malicious
edits and credential-related behavior~\cite{liu2026youragent}. We connect
several service properties within one sample: answer accuracy, content and
tool permissions, reported input, and account charges. Price grouping and
repeated measurements then show how the observed discrepancies are distributed
across services and whether they recur.

The workload matters when interpreting these findings. Xie et al. benchmark
methods for detecting unsafe outputs during generation and show that their
effectiveness depends on the task and application
scenario~\cite{xie2026onlinesafety}. Such detection evaluates output safety,
while identifying relay-added content also requires examining its source.
Controlled
man-in-the-middle experiments change inputs to induce factual-recall
failures~\cite{fastowski2025mitm}, while third-party plugin attacks manipulate
integration responses consumed by a model~\cite{zhao2024thirdparty}.
These studies explain how answers can change without model substitution and
why request handling and returned content need separate checks. Our benign,
standalone API requests may not trigger attacks that depend on agent tools,
credentials, or longer interactions. Consequently, the absence of confirmed
injection in our workload is compatible with attacks found in those settings.
The differences in workloads and samples also prevent pooling their rates
with ours or ranking detection methods by these counts.

Practitioner tools offer additional ways to monitor relays. Monitoring sites
publish rankings and run protocol, response-structure, and knowledge
checks~\cite{hvoy2026monitor,llmtest2026monitor}. The public descriptions
reviewed for this study do not provide the validation data and false-positive
calibration needed to interpret their scores as model-identity guarantees.
Open-source tools expose response, tool-call, and stream checks, although the
inspected project provides synthetic example reports without a comparable
service measurement sample~\cite{relayAudit2026tool}. An industry brief also
reports monitoring 171 endpoints over 14 days under its own detection and
repetition criteria~\cite{bazaarlink2026brief}. Our measurements use explicit
reference responses, decision thresholds, and reviewed labels to make the
basis of each reported discrepancy clear.

\section{Methodology}
\label{sec:methodology}

We test three aspects of relay service: whether answers differ from the
official model, whether returned content and tool calls respect the request,
and whether reported usage agrees with official counts and account charges.
The first two checks use matched workloads sent to relay and official APIs.
The third compares reported token quantities and account charges. Controlled
response mixtures calibrate the accuracy test, and a second round checks
whether the observed differences persist.

\subsection{Service Selection and Price Groups}
\label{sec:methodology:sampling}

We study 20 services offering GPT-6 Astra, Claude Opus 5, or both. We find
candidates through model recommendations, community resources, and web searches.
For model recommendations, we give ChatGPT, Claude, Gemini, and Grok the same
context and questions. Community resources include repositories and Hacker News
searches. Web discovery uses English search queries and recommendation lists.
We deduplicate domains and check each service's website, registration,
model catalog, interface, payment options, and prices.

\textbf{Eligibility and sample expansion.}
Services must permit individual registration and API access to a target model.
The first batch, selected through model recommendations, required both
targets. Subsequent
recruitment allowed either target and required an independent site, self-service
registration, metered access, public prices, an English interface, and payment
by credit card or PayPal with a minimum deposit of at most \$15.
All requests were sent from the USA, so results
describe service delivered to clients in one region.
These criteria exclude relays that require cryptocurrency payment, non-English
interfaces, enterprise registration, or large deposits, so the sample does not
represent the whole relay ecosystem.

The first batch contained only two low-price services. The sample was
expanded after a pilot of the four qualifying low-price services (batch 2),
whose continue-or-stop rule was fixed before any pilot response was viewed. During expansion, deduplicated recommendation lists and
web-search
results were interleaved in a fixed order. We checked candidates in that
order until each price group's recruitment target was filled, using the next
eligible service in the same group when a replacement was needed. All
qualifying low-price candidates were included up to a cap of eight, and the
same- or higher-price group was filled to about ten in the same order. The final
sample contains ten low-price and ten same- or higher-price services and is
not random.
Supplementary Section I describes recruitment, price classification, and
collection groups, which appear as batches in the result tables.

\textbf{Price classification.}
We compare standard short-context, non-cached input and output prices for
both targets. For service $s$, let $c_{s,j}$ and $c_{\mathrm{official},j}$ be
the relay and official prices for each of these four entries. The mean ratio is
\[
 P_s=\frac{1}{4}\sum_{j=1}^{4}
       \frac{c_{s,j}}{c_{\mathrm{official},j}}.
\]
We classify $P_s<0.95$ as low-price, $0.95\leq P_s\leq1.05$ as same-price,
and $P_s>1.05$ as higher-price. Published quotes remain in the calculation
even when a target cannot be measured through the required interface.

For services with multiple route groups, we use the new key's default group
or the recommended individual-developer group when a choice is mandatory.
Prices correspond to the group actually used, excluding paid-unlock routes.
Price-range and currency-conversion rules are given in Supplementary Section I.
The groups reflect advertised prices at verification.

\subsection{Question Selection}
\label{sec:methodology:probes}

Accuracy comparisons require questions that the declared model can answer
and plausible cheaper substitutes may miss. We therefore pre-validate a bank
of 96 candidate questions in three categories: recent-event knowledge,
parameter-scale-sensitive tasks, and capability-edge tasks. These categories
guide question design and do not isolate training cutoff or parameter count.
Each question has a prompt, answer key, scoring tolerance, and supporting source.

We sample each candidate three times on both official targets and three
cheaper substitutes: Claude Haiku 4.5, GPT-6 Luna, and GLM-4.5-Flash.
A question qualifies for a target if that target answers correctly in at
least two trials and at least two substitutes each fail in at least two
trials. From qualifying questions, we select 25 per target: nine knowledge,
eight parameter-scale, and eight capability-edge questions. The primary bank
prioritizes larger target--substitute contrasts within these quotas. An
alternative bank uses the original question order. We collect their union
of 27 questions per target to check whether findings depend on the selection rule.
The banks share 23 of 25 questions (92\%), so this checks local selection
sensitivity, not independent task coverage.

The bank, prompts, scorer, and selection rules were revised using official
and substitute pilot responses and frozen before formal relay collection
(Supplementary Section V). The contrast depends on the tested panel, and
untested substitutes may answer the entire bank. Because questions are
selected on the same pre-validation responses of these substitutes, the
mixture simulations below reuse that selection and may give an
optimistic view of detectability.

\subsection{Answer Collection and Accuracy Comparison}
\label{sec:methodology:a}

\textbf{Matched requests.}
Each relay--target pair and official reference receives five scheduled trials
per question, giving 135 calls for the 27-question union and at most 125 scored
responses for the primary bank. We request \texttt{low} generation effort
where supported, using the supported settings for each substitute.
Messages requests use the same no-search system instruction in both arms,
with an unused declared tool and \texttt{tool\_choice=none}. Each request is
standalone. Generated tool calls and commands are never executed.

\textbf{Scoring and missing responses.}
For each relay, target, question, and trial, we retain the first successfully
parsed HTTP 200 outcome. Retries cannot replace a completed answer or safety
refusal. Timeouts, protocol errors, empty unanswered responses, and unparseable
outcomes are treated as failed calls and do not count as incorrect answers.
A question enters the comparison only when both arms have a scored outcome.
Each arm then contributes all available trials for that question, without
requiring trial-level pairing. Safety refusals count as misses in the primary
analysis and are excluded in a sensitivity analysis.

The frozen scorer assigns binary credit using exact, substring, or numeric
matching where supported by the answer key and tolerance. Free-form answers
require review under the scoring rubric. Credit depends on the requested
answer content, with incomplete compound answers receiving zero.

\textbf{Answer review.}
\label{sec:methodology:annotation}
Free-form answers initially receive Claude Opus 5.5 labels. Manual review
sheets hide relay and target identity, official-reference status, and initial
judge labels. Identical answers to the same question reuse adjudicated labels.
The original collection uses one fully manual and one model-assisted review
workflow. Later collections use two fully manual workflows. Agreement is
measured before discussion, and disagreements and uncertain items are
adjudicated without changing the frozen bank.
Supplementary Section IV gives review procedures, agreement, and label-reuse
coverage.

\textbf{Detecting accuracy differences.}
For a relay and its official reference, let $h_r/n_r$ and $h_o/n_o$ be
their scored hit rates. The reported risk difference (RD) is
\[
 \mathrm{RD}=h_o/n_o-h_r/n_r.
\]
Positive RD means lower relay accuracy. We compare pooled hit counts using
a two-sided Fisher exact test and denote its Benjamini--Hochberg
(BH)-adjusted value by $q$~\cite{benjamini1995fdr}. The fixed rule labels a pair:
\begin{itemize}
 \item \emph{Confirmed behavioral inconsistency} if $q<0.05$ and
       $|\mathrm{RD}|\geq0.20$.
 \item \emph{Suspected behavioral inconsistency} if $q<0.05$ with a smaller
       effect, or $0.05\leq q<0.15$ with $|\mathrm{RD}|\geq0.20$.
 \item \emph{No evidence} otherwise.
\end{itemize}
These are operational labels under the stated rule. ``Confirmed'' denotes its
larger-effect branch, not independently verified substitution or persistent
service degradation. Neither label identifies the backend model.
Because the five trials of a question share errors, the Fisher test treats
correlated trials as independent; the question-block check below covers only
the originally flagged cells.
The two-sided rule also detects higher relay accuracy, which may reflect
retrieval or other processing.

The first-round analysis applies BH separately to each collection batch,
across all testable relay--target pairs and both targets within that batch.
This departs from the planned combined first-round family. The resulting
within-batch $q$ values do not provide whole-cohort false-discovery control. The separate native-client supplement uses a nominal exploratory
$p$ value outside these families.

\textbf{Price comparison and sensitivity checks.}
At the service level, we compare the two price groups using a two-sided Fisher
exact test. The primary comparison uses the primary bank with refusals scored
as misses and counts a service when at least one target receives either
inconsistency label. Four analysis specifications cross the
primary or alternative bank with refusals counted as misses or excluded.
A separate robustness comparison requires at least one target to receive
an inconsistency label under all four specifications. This is a robustness check added after the primary comparison did not reach
0.05, not the primary price test, and its smaller $p$ value should not be
read as stronger evidence.

Because repeated trials and shared references induce dependence, a post-hoc
check for the original cohort resamples questions as paired blocks, retaining
their available trials. Supplementary Section III gives the bootstrap settings.
Table~\ref{tab:question-intervals} reports intervals conditional on the selected
question bank. These three-cell intervals do not validate the full cohort.
Human scoring agreement does not establish trial independence; pooled $p/q$
values may be optimistic, and service counts summarize operational flags.

\subsection{Content and Tool-Permission Checks}
\label{sec:methodology:b}

\textbf{Screening and contextual review.}
To check for unwanted additions to otherwise benign responses, we use code
and setup requests in five classes: package names, hidden shell commands,
suspicious URLs, encoded content, and installation instructions. Five wording
variants and two trials per class give 50 scheduled responses per relay--target
pair. Both official targets receive the same workload.

Automated screens select responses containing package-name lookalikes,
unexpected domains, potentially harmful commands, or encoded content.
Supplementary Section V gives the matching rules and parser correction.

The authors then examine each candidate in context for task-irrelevant
packages, destinations, commands, installation sources, or executable encoded
additions. Task-related recommendations and parser errors are benign. Let
$r_r$ and $r_o$ be the proportions of successful relay and official responses
confirmed as injection. A finding requires both a positive excess $r_r-r_o$
and at least one confirmed injected response. Thus a screen match alone does
not count as injection.

The pooled response denominator includes eligible relay cells, excluding
official references, the separate native-client supplement, and an incomplete
cell. For zero confirmations, we report the one-sided 95\%
Clopper--Pearson upper limit under a binomial model. Correlated trials,
imperfect screening, and activation outside the tested workload limit its
scope. Supplementary Section V explains the scanner correction and its
unmeasured effect on candidate counts.

\textbf{Returned tool calls.}
We separately compare returned tool-call records with the permissions in
their requests, distinguishing explicit \texttt{tool\_choice: none} from
requests with no tool declaration. The count includes responses from failed attempts,
so it measures returned calls, not independent requests or executed actions.

\subsection{Usage and Charge Checks}
\label{sec:methodology:usage}

In the second round, we subtract official input-token counts from relay
counts for identical requests and summarize the differences by relay--target
pair using medians. For Messages responses, total input is
\texttt{input\_tokens} plus \texttt{cache\_creation\_input\_tokens} and
\texttt{cache\_read\_input\_tokens}. For chat completions, it is
\texttt{prompt\_tokens}; its \texttt{cached\_tokens} detail is already included.
When both formats occur, we use the latter fields, matching the saved parser.
We also record model names, reasoning details, and response metadata without
changing requests. These fields are controlled by the endpoint.

To determine whether reported usage enters the bill, we read account balances
immediately before and after collection for the six services exposing a
balance endpoint. We calculate the collection cost by multiplying each billed
token category's reported usage by its listed rate and summing across requests.
We then compare that total with the balance decrease. Dashboard snapshots
taken only after collection lack a baseline and are excluded. Agreement
shows how the reported usage is charged. It does not verify the tokens
processed upstream.

\subsection{Calibration with Official and Mixed Responses}
\label{sec:methodology:controls}

\textbf{Official configuration check.}
We test whether the official APIs support disabling reasoning by requesting
disabled thinking for Opus and no reasoning for Astra. Neither tested option
provides the intended control, so the planned full no-reasoning reference
collection was not performed. This is a configuration check, not a causal
test of reasoning effects. Supplementary Section II gives the calls.

\textbf{Controlled response mixtures.}
We test sensitivity to substitution by mixing collected official and substitute
answers to the same questions. One official collection supplies the comparison
reference, a later collection supplies the honest response pool, and
pre-validation supplies substitute responses. For each target, substitute,
and replacement probability $r\in\{0,0.10,0.25,0.50,1.00\}$, we generate
1,000 simulated pairs with five trials per primary-bank question.

Each trial samples a scored response with replacement from the substitute
pool with probability $r$, otherwise from the official pool. Missing substitute
pools fall back to the official pool. Thus $r$ is a sampling probability,
not an exact replacement count. Substitute pools contain at most three
responses per question.

Each simulation uses the accuracy comparison's Fisher test, effect threshold,
and label rules. It contains only one test, so no BH correction is applied and
the
nominal $p$ value is used in place of $q$. Detection frequency is the fraction
receiving either inconsistency label. At $r=0$, the comparison uses official
responses from two collections, providing an official-to-official control.
Supplementary Section II gives the response-pool dates and detection frequencies.

The mixture design was specified after the pilot and before simulation. It
reuses the selected bank and existing answer labels, with no new API calls or
independent
validation set. Its small response pools and single-test analysis mean the
frequencies cannot be read as field detection power or population false-positive
rates. Selection of both questions and substitutes can make these conditional
simulations optimistic; they are not upper bounds on detection rates.

\subsection{Repeated Measurements}
\label{sec:methodology:longitudinal}

To test whether the observed differences persist, we repeat measurements
11--18 days later with fresh same-day official references. We retain the same
questions, trial counts, requested effort, scoring rules, and thresholds.
The second round covers 35 of the 36 first-round relay--target pairs and uses
one BH family across its testable pairs, separate from the first-round families.
Partial collections follow the same missing-response rules, without bypassing
access restrictions or imputing unavailable responses.
Section~\ref{sec:results:rq4} explains the affected pairs and changes in
classification, which alone cannot establish a change in upstream model.
We ran no dedicated official-to-official retest, so the rate at which labels
flip without any relay change is unknown. The official reference hit rates
across collections give a partial measure of reference variation.
We also apply one BH family per round to the same 35 retested pairs as a
post-hoc sensitivity check, keeping the original results for comparison.

\section{Results}
\label{sec:results}
\label{sec:empirical}

We first evaluate the accuracy test with controlled response mixtures, then
compare relay accuracy across price groups and screen returned content.
Repeated measurements show which differences persist. Finally, tool-call,
usage, and balance comparisons connect answer behavior to delivery and billing.

\subsection{RQ1: Accuracy Tests for Model Substitution}
\label{sec:results:rq1}

\begin{figure}[t]
\centering
\includegraphics[width=\linewidth]{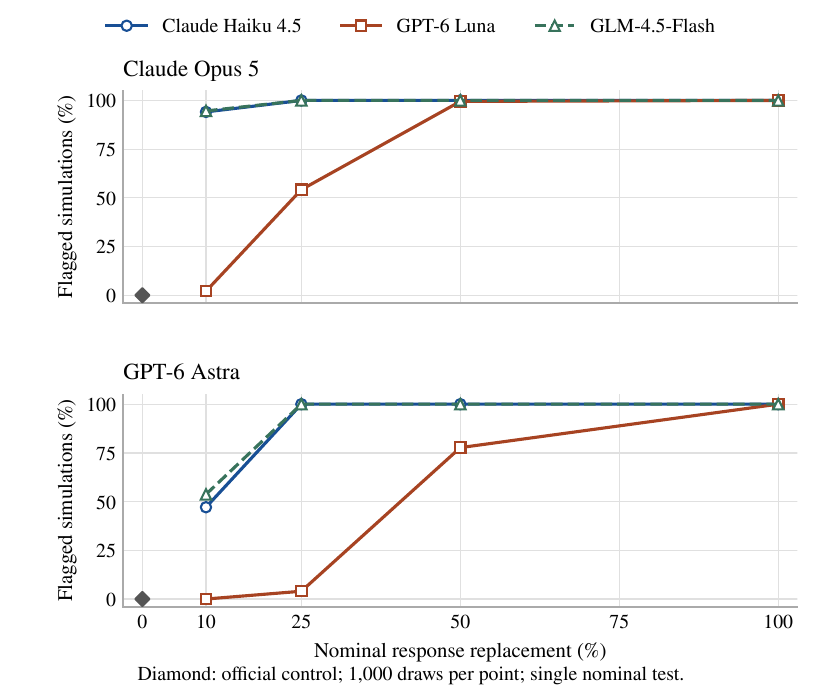}
\caption{Detection frequency by simulated replacement fraction, target model,
and substitute model.}
\Description{Two panels show detection frequency by replacement fraction for
Opus and Astra. Haiku and GLM reach complete detection at 25 percent,
while Luna requires larger fractions, particularly when replacing Astra.}
\label{fig:c2}
\end{figure}

\textbf{Weak substitutes are detected at moderate replacement fractions.}
Figure~\ref{fig:c2} and Supplementary Table S2 report the fraction of
simulated cells receiving either behavioral-inconsistency label. Each setting
uses 1,000 draws and one nominal test per cell. The zero-replacement control
is shared across substitutes within each target. For Haiku and GLM, every tested
25\%, 50\%, and 100\% replacement mixture is flagged for both targets.
At 10\%, the Opus comparisons flag 94.1\% for Haiku and 94.6\% for GLM,
while Astra flags 47.1\% and 53.7\%, respectively.
Detection at a fixed replacement fraction varies with the target and
substitute models.

\begin{table*}[t]
\caption{Answer accuracy and behavioral classifications in both rounds, grouped by
advertised price.}
\label{tab:exp-a}
\centering\footnotesize
\setlength{\tabcolsep}{2.5pt}
\begin{tabular*}{\textwidth}{@{\extracolsep{\fill}}llrrrrcrcrrrrcrc@{}}
\toprule
 & & \multicolumn{7}{c}{A} & \multicolumn{7}{c}{O} \\
\cmidrule(lr){3-9}\cmidrule(l){10-16}
 & & \multicolumn{5}{c}{Round 1} & \multicolumn{2}{c}{Round 2} & \multicolumn{5}{c}{Round 1} & \multicolumn{2}{c}{Round 2} \\
\cmidrule(lr){3-7}\cmidrule(lr){8-9}\cmidrule(lr){10-14}\cmidrule(l){15-16}
Relay & Batch & $r_r$ & $r_o$ & RD & $q$ & L & RD & L & $r_r$ & $r_o$ & RD & $q$ & L & RD & L \\
\midrule
\multicolumn{16}{@{}l}{\textit{Low price}} \\
R01 & 1 & 0.803 & 0.968 & +0.165 & \textbf{$<$0.001} & S & +0.272 & C & 0.704 & 0.992 & +0.288 & \textbf{$<$0.001} & C & +0.264 & C \\
R04 & 1 & -- & -- & -- & -- & -- & -- & -- & 0.912 & 0.992 & +0.080 & \textbf{0.023} & S & +0.056 & NNSS \\
R08 & 2 & 0.808 & 0.992 & +0.184 & \textbf{$<$0.001} & S & +0.008 & N & 0.840 & 0.992 & +0.152 & \textbf{$<$0.001} & S & +0.184 & SSCC \\
R11 & 2 & 0.744 & 0.992 & +0.248 & \textbf{$<$0.001} & C & +0.224 & CCSS & 0.418 & 0.992 & +0.574 & \textbf{$<$0.001} & C & +0.280 & C \\
R12 & 3 & 0.936 & 0.952 & +0.016 & 1.000 & N & +0.056 & N & 0.944 & 0.968 & +0.024 & 1.000 & N & +0.208 & CNCN \\
R13 & 3 & -- & -- & -- & -- & -- & -- & -- & 1.000 & 0.968 & $-$0.032 & 0.317 & N & -- & -- \\
R15 & 3 & 0.968 & 0.952 & $-$0.016 & 1.000 & N & +0.032 & N & 0.958 & 0.975 & +0.017 & 1.000 & N & +0.026 & N \\
R17 & 3 & 0.712 & 0.952 & +0.240 & \textbf{$<$0.001} & CCSS & +0.032 & N & 0.712 & 0.968 & +0.256 & \textbf{$<$0.001} & C & +0.248 & C \\
R19 & 2 & 0.936 & 0.992 & +0.056 & \textbf{0.041} & S & +0.000 & N & 0.976 & 0.992 & +0.016 & 0.368 & N & $-$0.024 & N \\
R20 & 2 & 0.704 & 0.992 & +0.288 & \textbf{$<$0.001} & C & +0.288 & C & 0.409 & 1.000 & +0.591 & \textbf{$<$0.001} & C & +0.583 & C \\
\midrule
\multicolumn{16}{@{}l}{\textit{Same price}} \\
R02 & 1 & 0.992 & 0.968 & $-$0.024 & 0.802 & N & +0.248 & CCSS & 0.976 & 0.992 & +0.016 & 0.809 & N & +0.024 & N \\
R03 & 1 & 0.960 & 0.968 & +0.008 & 1.000 & N & +0.016 & N & 0.976 & 0.992 & +0.016 & 0.809 & N & $-$0.016 & N \\
R05 & 1 & 0.976 & 0.968 & $-$0.008 & 1.000 & N & +0.024 & N & 0.984 & 0.992 & +0.008 & 0.809 & N & +0.045 & N \\
R06 & 1 & 0.944 & 0.968 & +0.024 & 0.809 & N & +0.040 & N & 0.952 & 0.992 & +0.040 & 0.389 & N & $-$0.008 & N \\
R09 & 3 & -- & -- & -- & -- & -- & -- & -- & 0.952 & 0.968 & +0.016 & 1.000 & N & +0.016 & N \\
R10 & 3 & -- & -- & -- & -- & -- & -- & -- & 0.776 & 0.968 & +0.192 & \textbf{$<$0.001} & S & +0.200 & C \\
R14 & 3 & 0.944 & 0.952 & +0.008 & 1.000 & N & +0.032 & N & 0.880 & 0.968 & +0.088 & \textbf{0.049} & SNNN & +0.000 & N \\
R16 & 3b & 0.936 & 0.960 & +0.024 & 0.571 & N & +0.056 & N & 0.968 & 0.992 & +0.024 & 0.571 & N & +0.024 & N \\
R18 & 3 & 0.960 & 0.952 & $-$0.008 & 1.000 & N & +0.040 & N & 0.976 & 0.968 & $-$0.008 & 1.000 & N & $-$0.024 & N \\
\midrule
\multicolumn{16}{@{}l}{\textit{Higher price}} \\
R07 & 1 & 0.960 & 0.968 & +0.008 & 1.000 & N & +0.016 & N & 0.960 & 0.992 & +0.032 & 0.554 & N & +0.000 & N \\
\bottomrule
\end{tabular*}
\par\smallskip
\begin{minipage}{\textwidth}\footnotesize
A: GPT-6 Astra; O: Claude Opus 5. $r_r$, $r_o$: relay and official hit rates;
RD: $r_o-r_r$. L lists four classifications in order: (1) primary bank, refusals
counted as misses; (2) primary bank, refusals excluded; (3) alternative bank,
refusals counted as misses; (4) alternative bank, refusals excluded.
One letter means all four agree; CCSS means C in (1)--(2) and S in (3)--(4).
C/S/N: Confirmed/Suspected/No evidence of behavioral inconsistency.
Hit rates, RD, and $q$ use the primary bank with refusals counted as misses;
only L compares all four specifications.
Bold: $q<0.05$; dash: outside the main cohort or unmeasured.
\end{minipage}

\end{table*}

\textbf{A related substitute can pass at a substantial replacement fraction.}
Astra-to-Luna mixtures are unflagged in all 10\% simulations and flagged
in only 4.0\% at 25\%. Detection rises to 77.7\% at 50\% and 100.0\%
at complete replacement. At complete replacement, only 12.9\% of Astra
simulations reach the larger-effect Confirmed branch, with 87.1\%
receiving Suspected. By comparison, Opus-to-Luna mixtures reach 54.2\%
detection at 25\% and 99.5\% at 50\%. The panel includes a model whose
recent knowledge can exceed Astra's, so the difficult contrast is
consistent with the bank's limited coverage of newer knowledge.

The difference between a flag and the larger-effect label matters.
For Haiku replacing Opus at 25\%, detection is 100.0\%, but only 77.7\%
of simulations meet Confirmed. For Haiku replacing Astra at 25\%,
52.6\% meet Confirmed and 47.4\% meet Suspected. These branches reflect
effect thresholds and significance, not a posterior probability that a
specific backend was served.

\textbf{The official-to-official control produces no flags.}
At zero replacement, official responses from the second batch are resampled
against the first batch's official reference. Each target has 0/1,000 flagged
simulations. This result assesses two reference collections, one selected bank,
and the available scored pools. The simulations repeatedly reuse those
responses and are not 1,000 independent production deployments. We do not
turn the empirical zero into a general false-positive bound.

\textbf{The configuration check does not isolate reasoning effects.}
The official calls did not implement the intended disabling settings:
Opus returned a thinking block despite the request, and Astra rejected
the no-reasoning value. The valid low-effort Astra call produced an answer
with reasoning-token usage. These observations support the configuration
description in Section~\ref{sec:methodology:controls}. They cannot rule out
relay-side changes to effort, instructions, or decoding.

Because the questions were selected on the same substitute responses that
supply the mixtures, these frequencies are optimistic. They characterize the
selected bank and substitute mixtures, not unseen models, adaptive probe recognition, or every implementation of
the advertised model.

\subsection{RQ2: Accuracy Differences from Official Models across Price Groups}
\label{sec:results:rq2}
\label{sec:empirical:results}
\label{sec:empirical:behavior}

\begin{figure}[t]
\centering
\includegraphics[width=\linewidth]{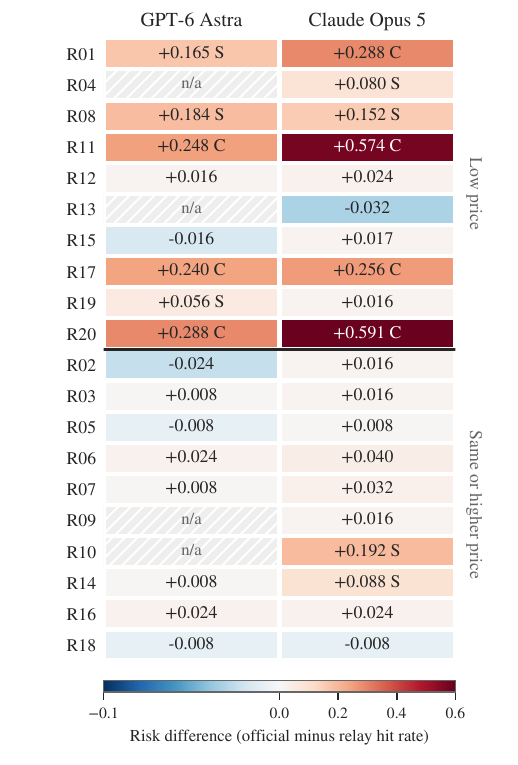}
\caption{First-round official-minus-relay accuracy differences for the 20 services,
grouped by advertised price. C/S: Confirmed/Suspected (primary analysis).
n/a: cells outside the main analysis.}
\label{fig:exp-a-heatmap}
\end{figure}

Table~\ref{tab:exp-a} reports the per-cell outcomes for all 20 measured
services. The primary analysis flags at least one cell at seven low-price
services and two same-price services. The remaining low-price services have
no flag, illustrating that price does not determine the result. An unflagged
cell does not exclude substitution, since Section~\ref{sec:results:rq1} shows
that a related substitute can pass at substantial replacement fractions. Labels refer
to behavioral inconsistency under the chosen prompts, references, and rules.
Figure~\ref{fig:exp-a-heatmap} summarizes the primary effects and missing cells.
Its hatching marks cells excluded from the main API analysis, and the
separately matched native-client comparison is not plotted.

Table~\ref{tab:exp-a} uses 125 scored relay responses per cell except, in
round~1, R01 A (122), R05 O (122), R11 O (122), R15 O (120), R19 O (123),
and R20 O (115), and in round~2, R05 O (76), R15 O (114), and R20 O (115).
First-round $q$ values retain per-batch BH families, including batch 3b's
separate official references. The second round uses one family.

\subsubsection{The Primary Price-Tier Test Gives p = 0.07}
\label{sec:empirical:price}

In the \textbf{primary analysis}, 7/10 low-price services have a flagged cell,
compared with 2/10 same- or higher-price services. The relay-level
two-sided Fisher test gives $p=0.0698$, reported as 0.07. This is the primary
price-tier test, and it does not meet a 0.05 threshold. The proportions suggest
an association among the selected services, not a statistically established
population price effect or a causal pricing mechanism.

\begin{figure}[t]
\centering
\includegraphics[width=\linewidth]{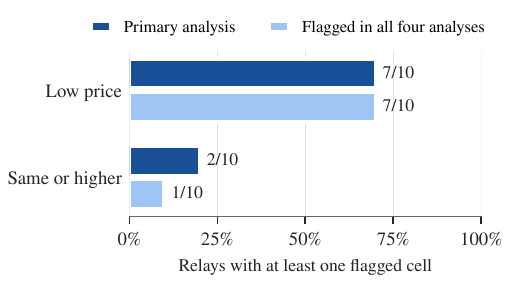}
\caption{Proportion of relays with behavioral differences by price group under the
primary and robustness analyses.}
\label{fig:price-tier}
\end{figure}

The \textbf{robustness analysis} counts a service only when at least one of
its cells is flagged in all four specifications. It gives 7/10 versus 1/10
and $p=0.0198$, reported as 0.02. That lower $p$ value comes from a specification added after the primary test
missed 0.05, so it is not the primary test and not stronger evidence.
Both comparisons remain conditional on the selected sample and the decision
to expand it after preliminary pilot results.
They also weight flags equally: R19 Astra's RD of 0.056 ($q=0.041$)
contributes one service, just as an RD above 0.5 does. Table~\ref{tab:exp-a}
retains effect sizes so that these unequal deficits remain visible.
Figure~\ref{fig:price-tier} keeps the two specifications separate.

\subsubsection{Large Deficits, Refusals, and Wrapper Effects}

\textbf{Broad deficits persist in several cells.} R01 Opus has a
hit rate of 0.704 against 0.992 officially, RD 0.288, and
$q<0.001$. Its failures span knowledge and capability tasks.
The pilot adds larger Opus deficits at R20 (0.409 versus 1.000,
RD 0.591) and R11 (0.418 versus 0.992, RD 0.574). Their Astra cells
are also flagged. Third-batch R17 is flagged for both targets, with
primary RD 0.240 for Astra and 0.256 for Opus. These deficits motivate
investigation, without identifying the backend weights or excluding changed
instructions, retrieval, or execution settings.

\textbf{The same-price R14 flag depends on refusal handling.} Its Opus
primary hit rate is 0.880 against 0.968, RD 0.088, and $q=0.0487$.
Only the primary bank with refusals scored as misses is flagged. Excluding
protocol-level refusals removes the flag, as does the alternative bank.
This cell accounts for the difference between the primary and all-four
service-level tallies. This pattern is refusal-sensitive behavioral
inconsistency. It does not identify a replacement model.

\textbf{The same-price R10 flag survives all four specifications.} Its
Opus rate is 0.776 against 0.968, RD 0.192, and $q<0.001$.
The service exposes a coding-client-style wrapper, with responses referring
to disabled tools. Those integration differences are relevant alternative
explanations. Persistence across scoring specifications cannot separate a
wrapper effect from model substitution. We therefore report behavioral
inconsistency and leave the causal backend claim unresolved.

\textbf{Not every classification level is invariant.} R17 Astra is
Confirmed in both primary-bank analyses and Suspected in both alternative-bank
analyses. It remains flagged in all four. This is the CCSS entry in Table~\ref{tab:exp-a}: changing the bank changes
the label from C to S, but all four analyses still detect a difference.
The first letter gives the primary result used in the price comparison.
The remaining letters show sensitivity to refusal handling and question
selection. A C-to-S change therefore preserves a flag, while a change to N
removes it. The robustness tally counts a cell only when no specification
returns N. The original three flagged API cells retain their labels across
all four specifications.

\subsubsection{Missing Cells and the Separate Client-Matched Supplement}
\label{sec:empirical:supplement}

The main table leaves four Astra cells missing. R04 requires an unmodified
first-party client and lies outside the original compatible API protocol.
R13 and R10 expose Astra only through the Responses interface,
which is excluded to keep the interface consistent with the official
chat-completions arm.
R09 changed key/group access during collection. Its partial responses
are excluded because no whole-cell collection under one consistent
configuration is available.
These cases leave the accuracy comparison unresolved.

R04's separate client-matched comparison for the first batch gives 117/125 against
123/125, RD 0.048, and nominal two-sided $p=0.1023$. Its alternative bank gives
119/125 against 123/125 and $p=0.2811$. Refusal exclusion does not change these
comparisons. The client-matched supplement does not
replace a missing main-cohort cell or enter the price-tier test as a new relay.

\subsubsection{Sensitivity to Repeated Answers}

Repeated answers to one question can share errors. The original-cohort
check therefore resamples the 25 shared questions as blocks, keeping each
question's official and relay trials together. Table~\ref{tab:question-intervals}
reports gaps that weight questions equally, with 95\% percentile intervals.
Missing calls explain why R01 Astra's question-mean gap is 0.184 while its
pooled-trial RD is 0.165. Supplementary Section III gives the resampling settings.
Grouping also preserves the pairing between the official and relay answers
to each question: a resampled question contributes both sides of the
comparison. The resulting intervals describe uncertainty across questions
within this bank, without treating repeated answers as new independent
questions.

All three intervals remain above zero, supporting the direction of these
accuracy deficits after grouping repeated answers. The intervals cover only
the original flagged cells, are conditional on the selected questions, and
are unadjusted for multiple comparisons. They do not resolve shared-reference
dependence or the separate batch-level correction families, and cannot
identify the cause of the accuracy gaps.

\begin{table}[t]
\caption{Question-level official-minus-relay accuracy gaps and bootstrap intervals.}
\label{tab:question-intervals}
\centering\small
\setlength{\tabcolsep}{4pt}
\begin{tabular}{@{}llrr@{}}
\toprule
Relay & Target & Mean gap & 95\% interval \\
\midrule
R01 & Claude Opus 5 & 0.288 & [0.128, 0.464] \\
R01 & GPT-6 Astra & 0.184 & [0.080, 0.304] \\
R04 & Claude Opus 5 & 0.080 & [0.016, 0.160] \\
\bottomrule
\end{tabular}
\end{table}

\subsection{RQ3: Injected Content in Code and Setup Responses}
\label{sec:results:rq3}
\label{sec:empirical:injection}

Table~\ref{tab:exp-b} summarizes the \textbf{1,686 relay-arm responses} from
the first-round code and setup workload and the candidates selected for review.
There are \textbf{zero confirmed
injections}. The one-sided 95\% Clopper--Pearson upper limit is 0.00178,
or \textbf{0.18\%}. This bound applies under the binomial model to confirmations
in the measured workload. It does not cover all malicious edits, account histories, or production workloads.

Two authors classify all 64 first-round candidates (48 relay-arm, 16
official-arm) across the four collections as benign. Legitimate task-related packages,
known example destinations, and parser errors account for reviewed detections.
The initial batch includes package-extraction false positives caused by shell
continuations and requirements-file arguments. No candidate is promoted to
an injection merely because the detector flagged it. Every measured cell's
confirmed rate and excess over its official reference are zero.

The original compatible-API workload contributes 546 relay responses. Its
separate client-matched supplement has 50 further relay responses and 50
official responses, all without candidates. These supplement responses are
not included in the 1,686 cohort denominator. Neither are official arms nor
the incomplete R09 Astra cell.

The second round adds 1,646 relay-arm responses with 43 automatic candidates.
Two authors classify 40 identically as benign. Three split labels on
optional scikit-learn suggestions for loading Iris are adjudicated benign,
matching seven earlier candidates of the same form, one from an official arm.
No round-2 injection is confirmed. Pooling both rounds gives \textbf{0/3,332
confirmed events}. Assuming independent binomial outcomes, the one-sided 95\%
upper limit for confirmations is 0.09\%; it is not a bound on actual malicious content.

\begin{table}[t]
\caption{First-round content-screening candidates and analyzed relay responses,
grouped by advertised price.}
\label{tab:exp-b}
\centering\small
\setlength{\tabcolsep}{4pt}
\begin{tabular}{@{}lcc@{\hspace{14pt}}lcc@{}}
\toprule
\multicolumn{3}{@{}c}{\textit{Low price}} & \multicolumn{3}{c@{}}{\textit{Same or higher price}} \\
\cmidrule(r{7pt}){1-3}\cmidrule(l{0pt}){4-6}
Relay & A & O & Relay & A & O \\
\midrule
R01 & 0/47 & 4/50 & R02 & 0/50 & 2/50 \\
R04 & --   & 4/49 & R03 & 0/50 & 5/50 \\
R08 & 0/50 & 3/50 & R05 & --   & --   \\
R11 & 0/50 & 0/42 & R06 & 0/50 & 5/50 \\
R12 & 0/50 & 2/50 & R09 & --   & 2/50 \\
R13 & --   & 1/50 & R10 & --   & 1/50 \\
R15 & 0/50 & 6/50 & R14 & 0/50 & 1/50 \\
R17 & 0/50 & 1/48 & R16 & 0/50 & 3/50 \\
R19 & 0/50 & 3/50 & R18 & 0/50 & 1/50 \\
R20 & 0/50 & 3/50 & R07$^\dagger$ & 0/50 & 1/50 \\
\midrule
Total & 0/397 & 27/489 & Total & 0/350 & 21/450 \\
\bottomrule
\end{tabular}
\par\smallskip
\begin{minipage}{\columnwidth}\footnotesize
Entries: relay-arm candidates/analyzed responses (the 16 official-arm
candidates are excluded). A: GPT-6 Astra; O: Claude Opus 5.
$^\dagger$: higher-price service; dash: unmeasured or excluded.
\end{minipage}

\end{table}

Repeated prompts and services need not generate independent Bernoulli outcomes.
The exact binomial interval is reported with that assumption, while unknown
screening sensitivity further limits its reach. The workload excludes executed
tools, secrets, and long histories that could activate conditional injection.
A zero confirmation is a bounded negative result under these conditions,
not certification that any service is safe. Without rerunning both parsers on the initial workload, we cannot quantify
how the correction changed candidate counts.

\subsection{RQ4: Accuracy Differences over Time}
\label{sec:results:rq4}

The second round ran 11--18 days after the first-round
collections (Section~\ref{sec:methodology:longitudinal}). Of the 36 first-round
main cells, 35 were retested. R13 Opus requires a named coding client.
Table~\ref{tab:exp-a} compares results across rounds.

\textbf{Six services are flagged in both rounds.} Of the 35 recollected cells,
9 are flagged in both rounds, 5 lose their primary flag, 2 gain one, and 19 have
no flag in either. At the service level, R01, R08, R10, R11, R17, and R20 are
flagged in both rounds. R04, R14, and R19 are no longer flagged, and R02 and R12
are newly flagged. The round-2 primary tier comparison gives 6/9 low-price
versus 2/10 same- or higher-price services, again $p=0.0698$, with R13
unmeasured. The all-four count gives 5/9 versus 2/10 ($p=0.17$).

The original seven flag transitions use different BH families across rounds.
With one 35-pair family in each round, six pairs gain or lose a flag: R19
Astra is no longer flagged in round~1 ($q=0.089$), while all other primary
labels are unchanged. This post-hoc check isolates correction-family effects,
while response dependence and reference variability remain. Supplementary
Section III gives the calculation; anonymous counts accompany the paper.

\begin{table*}[t]
\caption{Observed tool calls, response fields, reported usage, and account charges.}
\label{tab:delivery}
\centering\small
\setlength{\tabcolsep}{4pt}
\begin{tabular}{@{}>{\raggedright\arraybackslash}p{.27\textwidth}
>{\raggedright\arraybackslash}p{.28\textwidth}
>{\raggedright\arraybackslash}p{.40\textwidth}@{}}
\toprule
Observation & Services (count) & Supported conclusion and limit \\
\midrule
Tool calls despite suppression & Round 1, O: R04 (1), R11 (24), R10 (1), R17 (4); round 2: none &
Returned \texttt{noop} calls contradict \texttt{tool\_choice: none}; execution and
modification point are unknown. \\
Search calls without supplied tools & Round 1, A: R08 (2); round 2: none &
Returned \texttt{web\_search} calls; actual retrieval is unverified. \\
\midrule
Streaming for a non-streaming request & R19 &
Stream semantics differ; says nothing of backend identity. \\
\texttt{<think>} tags in output & R01 (A) &
Format difference, not a substitute fingerprint. \\
Cross-format refusal representation & R15 (O) &
\texttt{content\_filter} treated as a safety refusal, not a failed call. \\
\midrule
Key disabled after safety refusal & R05 (three keys, both rounds) &
Sequence observed; operator rationale unknown. \\
Client-restricted access & R13 (O, round 2) &
Requires a named coding client; not bypassed. \\
\midrule
Input-token surplus (round 2) & R01, R11, R20 (A, 4.38k, 4.22k cached); R08 (A, 4.22k; O, 734);
R10 (O, 59.5k); R16 (O, 274); R11 (O, content screening, 10.6k); R20 (O, content screening, 5.4k) &
Relay-reported usage: hidden input or altered accounting, not which. \\
Input-token deficit & R15 (O, $-$355), R20 (O, $-$275), R04 (O, $-$97) &
Relay-side token counting. \\
Dated model name for A & R05 (all), R12 (101/185) &
Two returned names for one request at R12; route unverified. \\
Balance vs.\ reported usage (round 2) & Equal at R03, R04, R11, R12, R17, R19 &
Charges follow reported usage at listed prices; others not reconcilable. \\
\bottomrule
\end{tabular}
\par\smallskip
\begin{minipage}{\textwidth}\footnotesize
A: GPT-6 Astra; O: Claude Opus 5.
\end{minipage}

\end{table*}

\textbf{Lost flags are mostly small effects.} Three of the five lost flags had
first-round RDs below 0.09 (R04 O, R14 O, R19 A), consistent with movement
around the threshold. Two larger Astra effects vanish: R08 from 0.184 to
0.008 and R17 from 0.240 to 0.032, while both services' Opus cells stay
flagged. Upstream routing, model updates, or account-level allocation could
produce such changes, which these measurements cannot distinguish.

\textbf{A same-price service changes.} R02 Astra moves from RD $-0.024$ to
$+0.248$. Its round-2 errors include statements that the 2025 Nobel Prize in
Literature and Super Bowl LX had not yet occurred, plus wrong counting and
arithmetic answers, while other trials of the same items are correct. The
pattern is consistent with part of the traffic reaching a model with an
earlier knowledge boundary, without identifying that model. R12 Opus is
flagged only when refusals count as misses: 29 of its 135 responses refuse
benign items, against one in the official arm, and no behavioral difference is detected when refusals are excluded.

\textbf{Two rounds bound, but do not characterize, change.} The short-interval
repeat of R01 Astra within the first batch (95/121, RD 0.183) matched its primary
result (98/122, RD 0.165) at short range, with shared questions, a shared
reference, and three backfilled keys. Official reference hit rates range
from 0.952 to 0.992 for Astra and from 0.968 to 1.000 for Opus across
collections. This observed variation is smaller than an RD of 0.20 but
comparable to effects near the Suspected boundary. Without a dedicated
official-to-official retest, the rate of flag changes without service changes
is unknown. Two rounds cannot distinguish service changes from reference
and response variation or estimate how often routes change, and a
stable label does not establish that every request used the same route.

\subsection{RQ5: Tool Permissions, Usage Reports, and Charges}
\label{sec:results:rq5}

Table~\ref{tab:delivery} separates systematically tallied tool-call records
from qualitative protocol, text, and account observations. Four services
return tool calls despite \texttt{tool\_choice: none}. The counts are
1 at R04, 24 at R11, 1 at R10, and 4 at R17. The returned
tool name is the unused declared \texttt{noop}. A fifth service, R08,
returns two \texttt{web\_search} call records when no tool declaration was
supplied. These counts include calls returned during failed attempts, not tool
executions or rates. No official-arm response in either round returned a
tool call, and 24 of the 30 relay-arm calls come from one service. Qualitative observations have no systematic count, and
unknown counts are not treated as zero. Token surpluses in
Table~\ref{tab:delivery} are second-round median differences from official
counts for identical requests in the accuracy workload, with the
content-screening workload identified separately.

\textbf{A returned tool call is not a completed action.} The client executed
none of these calls. The records cannot distinguish relay modification,
upstream parameter noncompliance, or output translation. These counts include failed attempts and cannot be read as rates among
independent scored responses.

\textbf{Protocol handling changes the measurement denominator.} R19
returned server-sent events to a non-streaming request. The parser handles
this format without changing request bytes, while unparseable outcomes are
excluded from accuracy scoring. Treating them as incorrect answers would
conflate compatibility with capability. Conversely, R15's translated
\texttt{content\_filter} response is a safety refusal and counts as a miss
in the primary analysis.

\textbf{Some apparent instruction clues have known confounders.} R04 Opus
mentions a request not to search. Our client supplied the same no-search
instruction to both relay and official Messages arms. The phrase alone
therefore does not show that the relay added instructions. Coding-client-style text
at R10 likewise warrants checking for added instructions. It does not
identify a replacement model. The table also states what account enforcement and reasoning-format
observations can and cannot establish.

\textbf{Reported usage reveals unexplained token quantities.} On the same Astra item,
R01 and R20 report 4,416 input tokens and R11 4,418, with 4,224 cached in
each, where the official API reports 36. R08 reports a 4,216-token cached
surplus. A fixed,
cached surplus of this size is consistent with an added system prompt, but
altered accounting can produce the same reports. R10 Opus
reports a median surplus of about 59,500 tokens, mostly cache reads, in line
with its coding-client wrapper. Smaller surpluses (R08 and R16 Opus) and
deficits (R15, R20, and R04 Opus) indicate relay-side additions or token counting.
All five services with surpluses above 4,000 tokens are flagged in both rounds,
but R08's Astra cell has a surplus without a round-2 flag. A reported surplus
thus need not coincide with lower measured accuracy.

\textbf{Billing follows the reported usage.} For the six services exposing a
balance endpoint, the balance change equals the cost implied by their reported
usage at listed prices to the cent. R11 is the only service shared by this six-service group and the five
services reporting more than 4,000 extra input tokens. Its extra reported
quantities therefore enter the bill, without establishing that corresponding
hidden text existed. Whether the other four services bill their surplus
cannot be determined. Both usage and balances are endpoint-controlled.
The balance comparisons use collection totals across billed token types,
as illustrated in Figure~\ref{fig:scenario}; Supplementary Table S3 gives
R11's token categories, rates, and reconciliation. Returned model names mostly match the request, with
aggregator prefixes. R12 returns a dated Astra snapshot name for
101 of 185 responses and the undated name for the rest. No round-2 response
contains a tool call, including R11, which returned 24 in round 1.
These findings coexist at the service level across workloads and rounds;
they do not establish a shared cause or co-occurrence within one request.

RQ5 detects permission and delivery differences, but does not assess retention,
credential misuse, software patches, or authentication of upstream requests.

\section{Discussion}
\label{sec:discussion}

\subsection{Price and Relay Reliability}

Low-price services account for most persistent accuracy differences in our
sample, which was expanded after a low-price pilot, but the results do not establish price as a reliable predictor of
service quality. The primary price comparison does not reach the 0.05
significance threshold. Selection was nonrandom, and retesting the same
services adds no independent sample. Lower prices can reflect efficiency,
promotions, or different account arrangements, as well as changes that reduce
answer quality. Conversely, matching the official price does not ensure that
a relay preserves the requested service. Price is therefore useful for
organizing comparisons, but choosing a higher-priced relay cannot replace
checking its answers, tool permissions, and billed usage.

\subsection{Accuracy Differences and Model Identity}

An accuracy difference establishes a mismatch in observed answer quality,
without identifying the model that produced it. Model substitution is one
possible cause, but changed instructions, refusal policies, reasoning settings,
retrieval, and caching can also change answers. Removing an explicit refusal
instruction in our early pilot coincided with Opus scores rising from 6/40
to 20/40 (Supplementary Section V). This suggests prompt sensitivity, without
isolating a causal effect, reinforcing the wrapper explanation.
The sensitivity checks help
separate some of these explanations: a difference that disappears when refusals
are excluded has a different interpretation from one that persists across
question banks and refusal rules. Even a persistent difference cannot isolate
model substitution from an undisclosed wrapper. Likewise, passing the test
only shows that the selected questions did not reveal a difference, and the
mixture results show that a related cheaper model can pass at moderate
replacement fractions. Services that evade checks or imitate fingerprints pose further
challenges~\cite{zhang2026ghostprint,nasery2025robust}. These results support
using accuracy tests to identify services needing further checks, while keeping
claims about backend identity separate from measured behavior.

\subsection{Recommendations for Relay Users}

Users should check billed quantities alongside advertised rates. For identical
requests, the official API supplies a reference for reported input tokens.
A surplus warrants checking for added instructions or different token counting,
without assuming that every reported token was processed upstream. Where
balances are available, comparing their decrease with the calculated charge
shows whether reported usage enters the bill. Calculate charges separately
for cached input and other token categories because their rates can differ.
Agreement with the balance confirms the charge calculation, but does not
validate the reported quantity.

Accuracy comparisons should also be repeated with fresh official responses.
Keeping the questions, request settings, and scoring rules fixed helps make
changes interpretable. When a difference appears, users should first repeat
the comparison under the same conditions before changing the configuration.
Retaining the settings and responses allows later checks of whether a change
concerns answer correctness, refusals, or another returned field. Public
monitoring services can help locate candidates, but their ratings cannot
replace a comparison under the application's own requirements~\cite{hvoy2026monitor,llmtest2026monitor}.

Tool permissions require enforcement in the client. An application should
reject calls that violate its requested permissions and inspect tool names,
arguments, commands, and destinations before execution. Our finding of returned
calls despite no confirmed injection shows why content screening alone is
insufficient. Plugin and tool-metadata attacks also involve integration paths
beyond standalone responses~\cite{zhao2024thirdparty,li2026mcp}. These checks
should remain separate when reporting results: correct answers do not establish
correct billing, and a plausible answer does not authorize a tool call.

\subsection{Requirements for Verifiable Relay Services}

Our measurements identify what users can check today and what remains hidden.
A returned model name or usage count is a statement from the endpoint, not
independent verification of processing. To close this gap, a service would need
to connect the client's request to the processing and charges it reports.
Useful records should cover the requested model and permissions, disclosed
additions to the input, returned tool calls, and billed token categories.
Users could then distinguish an agreed processing cost from an unexplained
charge and determine whether a tool restriction survived forwarding.
The record should also distinguish upstream output from changes made during
forwarding. Otherwise, a valid provider signature could leave later changes
and the final charge unexplained.

Provider signatures can authenticate covered fields, and attested forwarding
can constrain how participating infrastructure handles a
request~\cite{luo2026alignment,xie2026aegis,wang2026evidence}. Their value depends
on what they cover and which component the user trusts. Signing an answer alone
does not verify its billed usage, and signing a token count does not by itself
establish that the count was measured correctly. Such mechanisms require
support unavailable through the ordinary relay APIs we tested. They can
complement the behavioral, permission, and billing checks studied here.

\section{Ethics}
\label{sec:ethics}
\label{sec:ethics:artifacts}

Measurements use registered accounts and benign knowledge, reasoning, code,
and setup requests. Returned commands and tool suggestions are not executed.
Endpoint access restrictions are not bypassed. No confirmed injection
was found in the tested workload, so the study's injection-disclosure
criterion was not met.
We anonymize all measured services and retain the identity mapping privately
for confidential verification. The manuscript does not
name the measured services, their domains, or their route groups.
The accompanying data contain anonymized results and plot inputs for both
rounds, including accuracy, content screening, usage, and billing summaries.
Full requests, responses, failure logs, and annotations remain private because
they can expose service identities and account information. The supplied
anonymous hit counts and analysis script allow recomputation of the common-family
comparison. Other collection and scoring scripts and the complete frozen bank
are not supplied, so the summaries alone do not allow independent rescoring,
recalculation of every result, or reproduction of historical endpoint behavior.

\section{Conclusion}
\label{sec:conclusion}

We studied 20 LLM API relays across two rounds using official APIs as
references. The measurements found recurring accuracy differences, changes
between rounds, and returned tool calls that violated explicit restrictions
despite no confirmed injection in the screened responses. Usage and balance
checks also showed that extra reported input-token quantities enter R11's
bill, while their source remains unverified. These findings show why relay evaluation must connect answer
quality, request compliance, and billing: success on one check does not
establish that the full service meets the user's expectations. For users,
the practical implication is to verify token quantities as well as prices,
check returned tool calls before execution, and repeat comparisons as
services change. For researchers, the study provides a documented way to
relate these problems within the same services through public APIs and
account records. These checks make delivery problems observable without
internal access, while leaving their causes unresolved. More transparent
services would link requested instructions, reported usage, and charges to
verifiable records of processing, allowing users to assess whether they
received the service they paid for.

\section*{Acknowledgment}
OpenAI Codex~\cite{openai2026codex}
was used to polish the language in the abstract,
Impact Statement, main text, and supplementary material. The authors take
responsibility for the final manuscript.

\bibliographystyle{IEEEtran}
\bibliography{references}

\clearpage
\setcounter{section}{0}
\setcounter{subsection}{0}
\setcounter{table}{0}
\renewcommand{\thetable}{S\arabic{table}}
\renewcommand{\theHsection}{supp.\arabic{section}}
\renewcommand{\theHtable}{supp.\arabic{table}}
\twocolumn[\begin{center}
{\Large\bfseries Supplementary Material\par}
\vspace{0.5em}
{\large What the Middleman Delivers: Answer Behavior, Tool Permissions,\\
and Billed Usage in LLM API Relays\par}
\vspace{1.5em}
\end{center}]

\section{Service Selection and Collection}

Candidates were identified through model recommendations, community resources,
and English web searches. The initial model-recommendation search gave
ChatGPT, Claude, Gemini, and Grok the same context and two questions. After
domain deduplication, seven of 15 candidates qualified: two low-price, four
same-price, and one higher-price service. The exclusions were four enterprise-only services, one unavailable site,
two without target support, and one without API access.

The low-price pilot used community repositories, English web searches, and
Hacker News. Four services qualified, reducing the planned five-service pilot
and proportionally scaling its continuation threshold. Preliminary outcomes
triggered expansion through deduplicated, interleaved recommendation lists
and web-search results. The expansion added nine services: four low-price
and five same-price. Paid placements were excluded. Candidates were checked
in order against the stated access, payment, and deposit requirements,
with replacements drawn from the same price group.

\textbf{Collection groups and timing.}
Table~\ref{tab:supp-collection} identifies the groups used in the main result
table. The question bank and selection rules were fixed before the initial
relay collection. Batch 1 includes next-day responses compared with the
October 2 reference; it is not entirely a same-day comparison.
Subsequent groups used same-day official references. The second round
retained the questions and analysis rules with fresh official responses.

\begin{table}[t]
\caption{Collection dates and official references in October 2026.}
\label{tab:supp-collection}
\centering\small
\setlength{\tabcolsep}{5pt}
\begin{tabular}{@{}llll@{}}
\toprule
Group & Relay dates & Reference date & Services \\
\midrule
Batch 1 & Oct.\ 2--3 & Oct.\ 2 & 7 \\
Batch 2 & Oct.\ 5 & Oct.\ 5 & 4 \\
Batch 3 & Oct.\ 7 & Oct.\ 7 & 8 \\
Batch 3b & Oct.\ 9 & Oct.\ 9 & 1 \\
Round 2 & Oct.\ 20 & Oct.\ 20 & 19 \\
\bottomrule
\end{tabular}
\end{table}

The initial API analysis uses a Benjamini--Hochberg (BH) family of 13
relay--target pairs. Later first-round groups each use a separate family,
departing from the planned combined first-round correction. Round 2 uses
one family across its 35 testable pairs. The separate native-client comparison
uses its own official reference and nominal exploratory $p$-values.

\textbf{Price and route selection.}
Prices correspond to the key's default route group, or the recommended
individual-developer group when a choice is required. Paid-unlock routes
are excluded. For quoted price intervals, the lower bound for the comparable
short-context setting is used. The four-quote mean ratio in the main methods
allows a 5\% margin around the official price for the same-price group.
This rule was set before expansion responses were collected, without changing
the initial and pilot assignments.

Foreign-currency prices use the last available official midpoint exchange
rate. Route-group and currency checks were completed after
first-round scoring but before the cross-group price comparison. The affected
service remained in the same-price group. For the second round, the billed
route group was checked before scoring. R17's group charged half the official
rate for both targets, retaining its low-price classification.

\section{Calibration Details}

\textbf{Official configuration check.}
The second-batch calls requested disabled thinking for Opus and no reasoning
for Astra on a multiplication prompt. Opus accepted the request but returned
a thinking block and reported 182 thinking tokens. Astra rejected the
unsupported no-reasoning value with HTTP 400. The study's low setting returned
an answer with reported reasoning tokens. Neither option disabled reasoning, so the full no-reasoning reference
collection was not performed.

\textbf{Controlled response mixtures.}
Table~\ref{tab:supp-c2} reports detection frequencies for the tested replacement fractions.
Each percentage counts Confirmed or Suspected behavioral inconsistency
among 1,000 resampled cells under a single nominal Fisher test, without
multi-cell adjustment.
The reference uses the first batch's official responses, and the honest
response pool uses the second batch's official responses. Substitute responses
reuse pre-validation.
A missing substitute pool falls back to the official pool. The nominal fraction
is a sampling probability, not a fixed number of replacements in each cell.
These frequencies apply only to the selected response pools.

\begin{table}[t]
\caption{Detection percentages from 1,000 simulated cells per replacement setting.}
\label{tab:supp-c2}
\centering\small
\setlength{\tabcolsep}{5pt}
\begin{tabular}{@{}llrrrr@{}}
\toprule
Target & Substitute & 10\% & 25\% & 50\% & 100\% \\
\midrule
O & Haiku 4.5 & 94.1 & 100.0 & 100.0 & 100.0 \\
O & Luna & 2.1 & 54.2 & 99.5 & 100.0 \\
O & GLM-4.5-Flash & 94.6 & 100.0 & 100.0 & 100.0 \\
\midrule
A & Haiku 4.5 & 47.1 & 100.0 & 100.0 & 100.0 \\
A & Luna & 0.0 & 4.0 & 77.7 & 100.0 \\
A & GLM-4.5-Flash & 53.7 & 100.0 & 100.0 & 100.0 \\
\bottomrule
\end{tabular}
\par\smallskip
\begin{minipage}{\columnwidth}\footnotesize
A: GPT-6 Astra; O: Claude Opus 5.
\end{minipage}

\end{table}
\section{Question-Level and Common-Family Checks}

The original-cohort check resamples the 25 shared questions as blocks,
keeping available official and relay trials within each question. It uses
20,000 bootstrap draws with seed 20261005 and reports unadjusted 95\% percentile
intervals. Questions receive equal weight, including when their trial counts
differ because of missing calls. RQ2 reports the three originally flagged cells' intervals, which remain
conditional on the selected question bank.

\textbf{Common-family sensitivity check.}
For the 35 pairs measured in both rounds, we recompute two-sided Fisher
$p$-values from the primary-bank hit counts and apply BH separately to each
round's same 35 pairs. The RD and label thresholds are unchanged. Six pairs
gain or lose a flag, compared with seven under the original families.
Only R19 Astra's first-round label changes, from S to N ($q=0.0893$).
The accompanying anonymous counts and script reproduce this post-hoc check;
it retains pooled-trial dependence and is not a new reliability validation.

\section{Annotation Agreement and Label Reuse}

\textbf{Per-batch agreement.} Agreement is computed before discussion on blind sheets that hide relay,
target, official-reference status, and initial labels.
Reasoning blocks are removed, and identical answers are grouped by question
after case and whitespace normalization. Final adjudicated labels are stored
separately from the submitted labels.

\textbf{Batch 1.} One author labels fully manually and the other manually
checks model prelabels item by item. The sheets contain 153 distinct field
answers covering 524 API labels and
33 short-repeat labels, plus 100 sampled pre-validation answers. Of 253 items,
252 have two binary labels and 245 agree (97.22\%, Cohen's $\kappa=0.943$).
Seven disagreements and one uncertainty receive author adjudication. The
field/repeat subset has $\kappa=0.897$ and the pre-validation sample
$\kappa=0.940$.

\textbf{Batch 2.} Both authors annotate fully manually. Among 105 distinct
answers, 103 have two definite labels and 102 agree, with $\kappa=0.981$.
One disagreement and two uncertain items are adjudicated using the same
compound-answer and initializer rules as batch 1.

\textbf{Batch 3.} Both authors annotate fully manually. The main sheet
contains 108 items, including three
uncertain items. All 105 items with two definite labels agree ($\kappa=1.0$).
The uncertain items are adjudicated. A later six-item
supplement also has complete agreement. Its single-label distribution makes
$\kappa$ undefined, not 1.0. The batch-3b sheet contains 16 distinct
answers. Another 120 labels reuse identical, already adjudicated answers.
All 16 items agree ($\kappa=1.0$).

\textbf{Round 2.} Both authors annotate fully manually. The sheet contains 234
distinct answers. Another 1,011 labels reuse identical adjudicated answers.
Of 233 items with two labels, 228 agree ($\kappa=0.956$). Five disagreements
and one missing label are adjudicated under the batch-1 compound-answer,
initializer, and hedging rules. Both authors classify 40 of 43 injection candidates as benign. The three
disagreements are also resolved as benign after review.

Across first-round batches, both authors classify all 64 response-screening
candidates as benign. This count includes official-arm candidates. There is complete
observed agreement but no category variation, so injection-review $\kappa$
is undefined. Shared rubric errors remain possible.

\textbf{Label reuse.}
Reviewed-answer matches cover 174 of 480 valid non-automatic pre-validation
labels. The remaining 306 retain initial model labels. Of 70 free-form
native-client supplement answers, 59 match reviewed answers, two apply the
clarified Go initializer rule, and nine retain initial labels. Matching
reviewed answers and applying a rubric do not constitute direct human review
of every response. These later decisions do not reselect the frozen bank.
Within each target's primary bank, 84 non-automatic pre-validation labels
inform target or substitute scores. Reviewed-answer matches cover 42 for
Astra and 49 for Opus. Of the 63 non-automatic substitute-pool labels per
target used in the mixtures, 23 and 30, respectively, match reviewed answers.
The remaining labels retain initial model scores; automatic scores are
excluded from these denominators. Label transfer changed none of the valid
pre-validation scores, so it did not change frozen eligibility. This does
not independently validate the unmatched labels.

\section{Question Selection and Response Processing}

\textbf{Question-bank revisions.}
Before formal relay collection, official and substitute pilot responses led
to three changes: replacing DeepSeek Flash, whose initial rough matching
scored 30/40 recent-event answers, with GLM-4.5-Flash (0/40 on those items),
removing an instruction to answer unknown
questions with an explicit refusal, and adjusting automated scoring to ignore
Markdown emphasis and accept a final answer line. The refusal instruction
produced 6/40 correct Opus answers in an early pilot, compared with 20/40
after removal. The bank and selection rules were fixed before formal relay
collection.
Luna's documented knowledge cutoff is later than Astra's, so recent-event
questions alone cannot distinguish them. Capability questions must do so.

\textbf{Request handling and scoring.}
An early endpoint observation of unexpected search-tool availability led to
the unused declared tool and explicit tool suppression in Messages requests.
Both relay and official arms receive the same no-search instruction, so an
answer mentioning that instruction alone does not establish relay modification.
The response parser supports server-sent events even when the request
disables streaming. Parser corrections did not change request bytes. Outcomes
that could not be parsed are excluded from answer scoring. Complete correct
answers receive credit despite
reservations about verification, while incomplete compound answers receive
zero. The clarified Go initializer rubric accepts paraphrases
that convey the intended initial value without explicitly claiming an
incorrect two-argument syntax.
Second-round metadata collection records identifier prefixes, fingerprints,
and selected headers without changing request bytes.

\textbf{Content screening and parser correction.}
The package screen compares imports and installation names with expected
libraries and marks lookalikes at similarity at least 0.8. Other screens
inspect execution and dynamic-evaluation constructs, domains outside a
whitelist with subdomain-boundary checks, decodable encodings, custom package
indexes, and download-to-shell patterns. Encoding checks include prose and
comments. The other screens inspect code blocks.

The initial package scanner falsely treated shell continuation characters
and requirements-file arguments as package names. The corrected parser was
used for subsequent collections without changing requests. All candidates
from both scanner versions received contextual review. Because the initial
workload was not rescanned with both versions, the effect of the correction
on candidate counts cannot be quantified.

\section{R11 Usage and Charge Reconciliation}

Table~\ref{tab:supp-billing} includes both workloads and recorded attempts
with metadata, matching the original collection-total calculation. Ordinary
input excludes cache reads and writes. For each model, the calculated cost is
$(T_i p_i+T_r p_r+T_w p_w+T_o p_o)/10^6$; token categories and rates are
listed separately to avoid counting cached input twice. The total is
\$1.20107395, against a recorded deduction of \$1.2011, a difference of
\$0.00002605 at the stored precision. This reconciles endpoint-reported
quantities and charges, not upstream processing or undisclosed text.

\begin{table}[!ht]
\caption{R11 round-2 reported tokens and rates in USD per million tokens.}
\label{tab:supp-billing}
\centering\small
\setlength{\tabcolsep}{4pt}
\begin{tabular}{@{}lrrrr@{}}
\toprule
 & \multicolumn{2}{c}{Astra (188 calls)} & \multicolumn{2}{c}{Opus (185 calls)} \\
\cmidrule(lr){2-3}\cmidrule(l){4-5}
Category & Tokens & Rate & Tokens & Rate \\
\midrule
Ordinary input & 63,575 & 1.50 & 180,067 & 1.25 \\
Cache read & 756,388 & 0.15 & 722,046 & 0.125 \\
Cache write & 0 & 1.875 & 0 & 1.5625 \\
Output & 25,011 & 7.50 & 78,293 & 6.25 \\
\midrule
Cost (USD) & \multicolumn{2}{c}{0.39640320} & \multicolumn{2}{c}{0.80467075} \\
\bottomrule
\end{tabular}
\end{table}

\end{document}